\documentclass[]{aastex631}
\begin{document}

\title{Binding energies of small interstellar molecules on neutral and charged amorphous solid water surfaces.}

\author[0000-0002-4724-2782]{Tobe Vorsselmans}
\affiliation{Research group MOSAIC, Department of Chemistry, University of Antwerp, Universiteitsplein 1, B-2610 Antwerp, Belgium}
\author[0000-0002-3360-3196]{Erik C. Neyts}
\affiliation{Research group MOSAIC, Department of Chemistry, University of Antwerp, Universiteitsplein 1, B-2610 Antwerp, Belgium}

\begin{abstract}
The interstellar medium (ISM) is all but empty. To date, more than 300 molecules have already been discovered. Because of the extremely low temperature, the gas-phase chemistry is dominated by barrierless exothermic reactions of radicals and ions. However, several abundant molecules and organic molecules cannot be produced efficiently by gas-phase reactions. To explain the existence of such molecules in the ISM, gas-surface interactions between small molecules and dust particles covered with amorphous solid water (ASW) mantles must be considered. In general, surface processes such as adsorption, diffusion, desorption, and chemical reactions can be linked to the binding energy of molecules to the surface. Hence, a lot of studies have been performed to identify the binding energies of interstellar molecules on ASW surfaces. Cosmic radiation and free electrons may induce a negative charge on the dust particles, and the binding energies may be affected by this charge. In this study, we calculate the binding energies of CO, CH$_4$, and NH$_3$, on neutral and charged ASW surfaces using DFT calculations. Our results indicate that CO can interact with the surface charge, increasing its binding energy. In contrast, the binding energy of CH$_4$ remains unchanged in the presence of surface charge, and that of NH$_3$ typically decreases.
\end{abstract}


\section{Introduction} \label{sec:intro}

\begin{figure}[htb!]
\epsscale{0.50}
\plotone{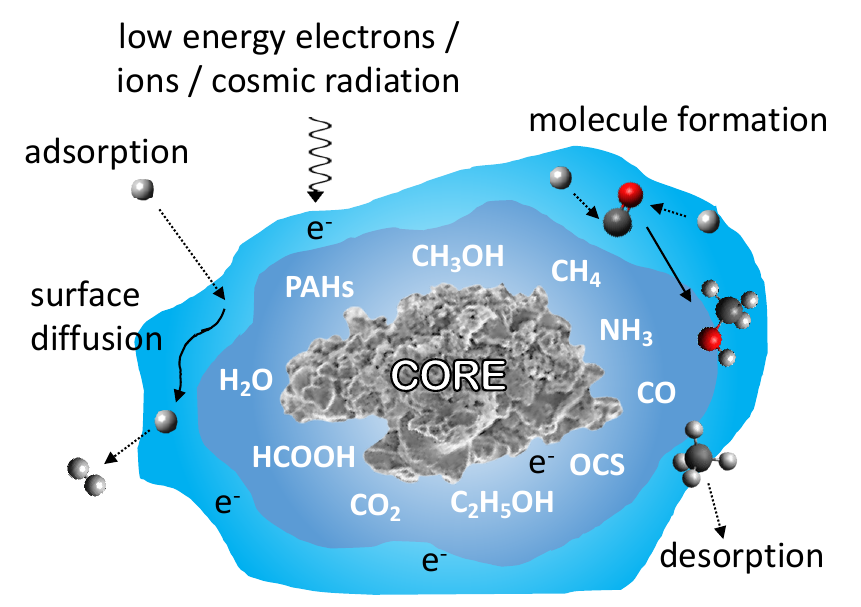}
\caption{Schematic representation of dust grains in the ISM.\label{fig:ice_mantle}}
\end{figure}

Considerable effort has already been made by the astrochemical community to study the binding energies on so-called ‘dust grains’ \citep{Duflot2021,Das2018,Wakelam2017,Tielens2005,Al-Halabi2004,Ferrero2020}. Figure \ref{fig:ice_mantle} shows a dust grain surrounded by an ice mantle infused with different molecules and the different processes that may happen on such grains. In molecular clouds, the core of the dust grain is surrounded by a layer of amorphous ice, also referred to as amorphous solid water (ASW). ASW can aid the reaction between interstellar molecules in three ways. Firstly, it can act as a passive third body that absorbs the excess energy released by a chemical reaction on its surface \citep{Pantaleone2020}. Secondly, it can also directly participate in the reaction by reducing the reaction barriers \citep{Enrique-Romero2019}. Thirdly, it can serve as a reactant concentrator by bringing molecules together at its surface \citep{Rimola2014}. These processes occurring on the grain surface are important to explain the observed abundances of molecules.

Interactions between gas-phase atoms/molecules and the dust grains are largely dependent on adsorption, diffusion, and desorption. An important variable for these processes is the binding energy of a molecule on the ASW mantle \citep{cuppen2017}. The binding energy will determine how likely it is for a molecule to stick to the surface, how easily it diffuses over the surface, and how easily it desorbs into the gas phase. These processes all contribute to the synthesis of new molecules in the ISM. A better understanding of these mechanisms thus provides a clearer insight into the evolution of molecular clouds and star formation \citep{Wakelam2017}.

A recent review discusses the thermal desorption of interstellar ices, explaining extensively the different methods used to find the binding energy of a molecule on interstellar ASW \citep{Minissale2022}. So far, most studies in the field of computational astrochemistry have only focused on first-order atomic or molecular desorption. Binding energies are, however, very much dependent on the chemical composition and morphology of the surface. Additionally, it has been shown that ASW is very likely to capture low-energy electrons. The number density of free electrons in molecular clouds is estimated to be $10^{-4}$ cm$^{-3}$ \citep{Rimola2021}. A study by Draine and Sutin from 1987 already recognized that grains in the interstellar medium can be negatively charged \citep{draine1987}. Rimola et al. performed a study on the binding energy of HCO$^+$ on a negatively charged ASW surface \citep{Rimola2021}. They found that cationic species in the ISM can react with thermalized low-energy electrons from an ASW surface through electron transfer to create the corresponding neutral radical.

Two recent papers examined the behavior of the hydroxyl radical on an ASW surface, which can become $^-$OH by electron attachment \citep{Tsuge2021, Woon2023}. Both mention that an extra electron on the surface will easily react with a surface bound radical creating an anion. This suggests that loosely bound electrons are a rare phenomena in the ISM, although the newly created anion can react with a cation to form a neutral species. According to these papers, $^-$OH is readily transported through the bulk of the ASW surface via a proton-hole transfer mechanism, increasing the likelihood of anion-cation recombination. Due to the high mobility of these anions and the presence of cations like H$^+$, an electron can remain weakly attached to the surface rather than binding with a radical. In this study we will focus on these loosely bound electrons and how they interact with adsorbing molecules, even though it is suggested that these would be rather rare in the ISM.

This work aims to examine the possible effect a charged surface has on the binding energies. To study this effect we use density functional theory (DFT) to first calculate the binding energies of CO, CH$_4$, and NH$_3$ on a neutral ASW surface. These binding energies are compared to the literature to evaluate our model. Subsequently, we charge the surface and recalculate the binding energies.

Our results indicate that CO can interact with the surface charge, increasing its binding energy. In contrast, the binding energy of CH$_4$ remains unchanged in the presence of surface charge, and that of NH$_3$ typically decreases.

\section{Method} \label{sec:method}

ASW surfaces are created by randomly placing 33 molecules in a cubic box with lateral dimensions of 10 {\AA} x 10 {\AA} x 10 {\AA}, using Packmol \citep{Martinez2009}. Due to this randomness, an amorphous cluster is guaranteed after optimization.

CO, CH$_4$, and NH$_3$ are positioned on 25 different locations, according to a 5x5 grid parallel to the xy-plane above the surface (figure \ref{fig:ASW_slab}d). The molecule is always placed 2 \AA above the surface along the z-axis, this ensures that the molecule interacts with the surface at the given x,y-coordinate. Because of the amorphous nature of the surface, a sufficiently large number of samples is required. Indeed, every site yields a different binding energy and a sufficient number of samples is required to allow us to map the interactions between the molecules and the ASW surface accurately.

Our calculations were performed using Gaussian 16 \citep{g16}. We used the hybrid PBE0 \citep{Perdew1996,Adamo1999} functional to approximate the exchange-correlation functional. Furthermore, we chose the split-valence triple zeta 6-311++G(d,p) basis set \citep{Weigend2006}. Adding the diffuse and polarization functions is important to accommodate the extra electron that will be added when the system is charged. Additionally, to correct for the inability of Kohn-Sham DFT to include dispersion interactions, the Grimme DFT-D3(BJ) dispersion correction was added \citep{Grimme2011}. We benchmarked the PBE0/6-311++G(d,p) method against the higher theory CCSD(T)/Aug-CC-pVTZ method, resulting in an error of only 10\%. The results of this benchmark is shown in table \ref{tab:Benchmark}. Additionally, previous research already applied the PBE-D scheme successfully to the CO/ASW system \citep{zamirri2018}. The CO system has the most delicate electronic structure of the three molecules, so it is within reason to adopt this scheme for the other systems as well.

\begin{deluxetable}{c c c c c c c}[hbt!]
    \tablecaption{Benchmark results for the binding energy of a water molecule in a small charged water cluster. \label{tab:Benchmark}}
    \tablehead{\colhead{Method} & \colhead{(H$_2$O)$_4^-$ (Ha)} & \colhead{(H$_2$O)$_3^-$ (Ha)} & \colhead{H$_2$O (Ha)} & \colhead{BE (kJ/mol)} & \colhead{$\Delta$ (kJ/mol)} & \colhead{MUE (\%)}} 
    \startdata
        CCSD(T)/Aug-CC-pVTZ & -305.39 & -229.03 & -76.34 & 44.39 & & \\
        PBE0/6-311++G(d,p) & -305.52 & -229.13 & -76.37 & 48.96 & 4.57 & 10\\
    \enddata
\end{deluxetable}

Adding a charge to the system is accomplished by changing the charge and spin multiplicity in the molecule specification section of the Gaussian 16 input file. However, we have to be careful with the interpretation of the results due to the sensitivity of DFT on charged systems. We therefore carefully checked our method against the results of Rimola et al., i.e., we calculated the binding energy of the HCO$^+$ cation on the negatively charged ASW using our computational settings and compared the results with the results reported by Rimola et al.\citep{Rimola2021}

The binding energy of a molecule to a surface is calculated using equation \ref{eq:binding}. This requires fully optimized energies for three different systems: the molecule, the ASW surface, and the molecule bonded to the surface. A zero point energy correction is often employed to this binding energy to obtain the energy of the ground state. In this study, we are mainly concerned with the change in binding energy caused by the addition of surface charge. As a result, we did not calculate zero-point energy correction for every configuration to reduce the computational cost. However, we did calculate the zero point energy of every unique end configuration. This made it possible to verify the end configuration for a true minimum and we can additionally estimate the zero point energy correction via extrapolation from the calculated zero point energies.

\begin{equation} \label{eq:binding}
    BE = [E(molecule) + E(ASW)] - E(molecule + ASW)
\end{equation}

\section{Results} \label{sec:results}

\begin{figure}[htb!]
\gridline{\leftfig{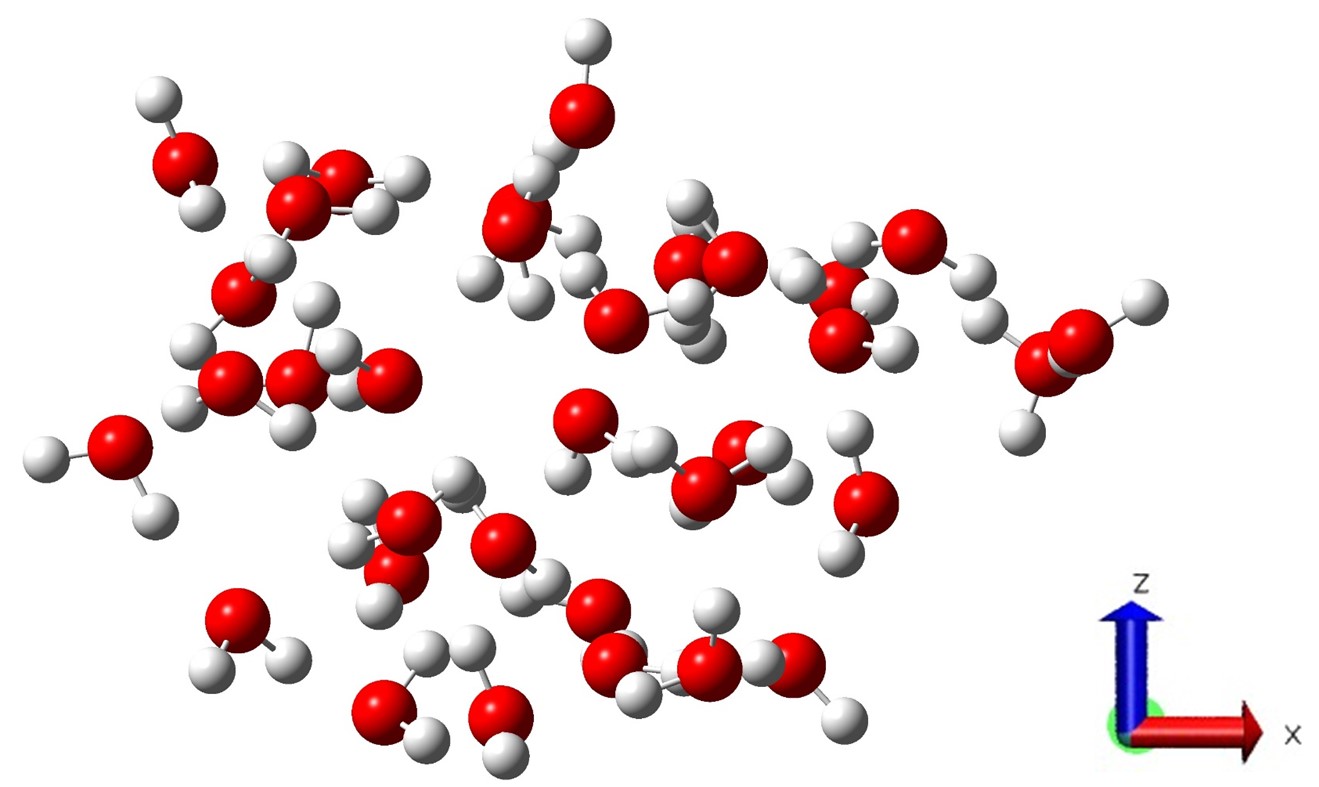}{0.4\textwidth}{(a) Neutral ASW}           
          \rightfig{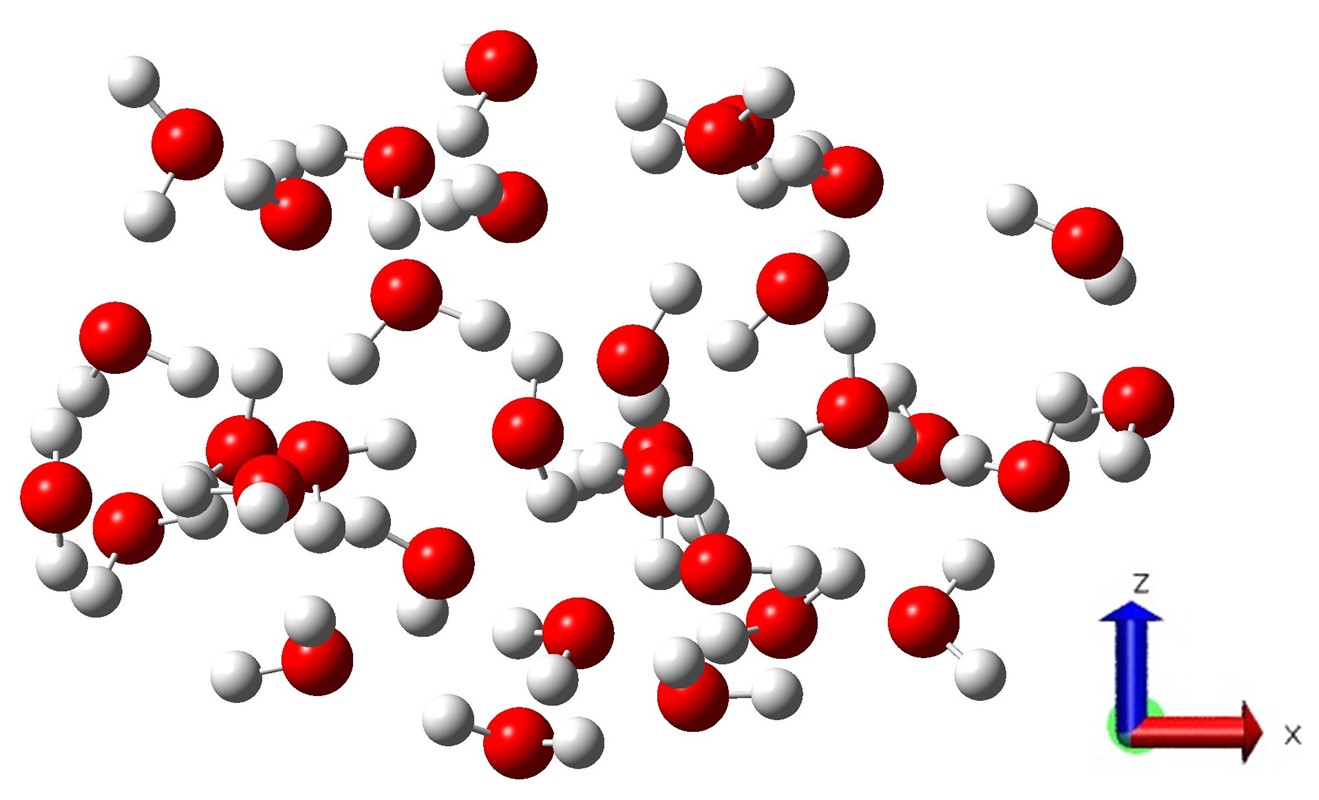}{0.4\textwidth}{(b) Charged ASW}}
\gridline{\leftfig{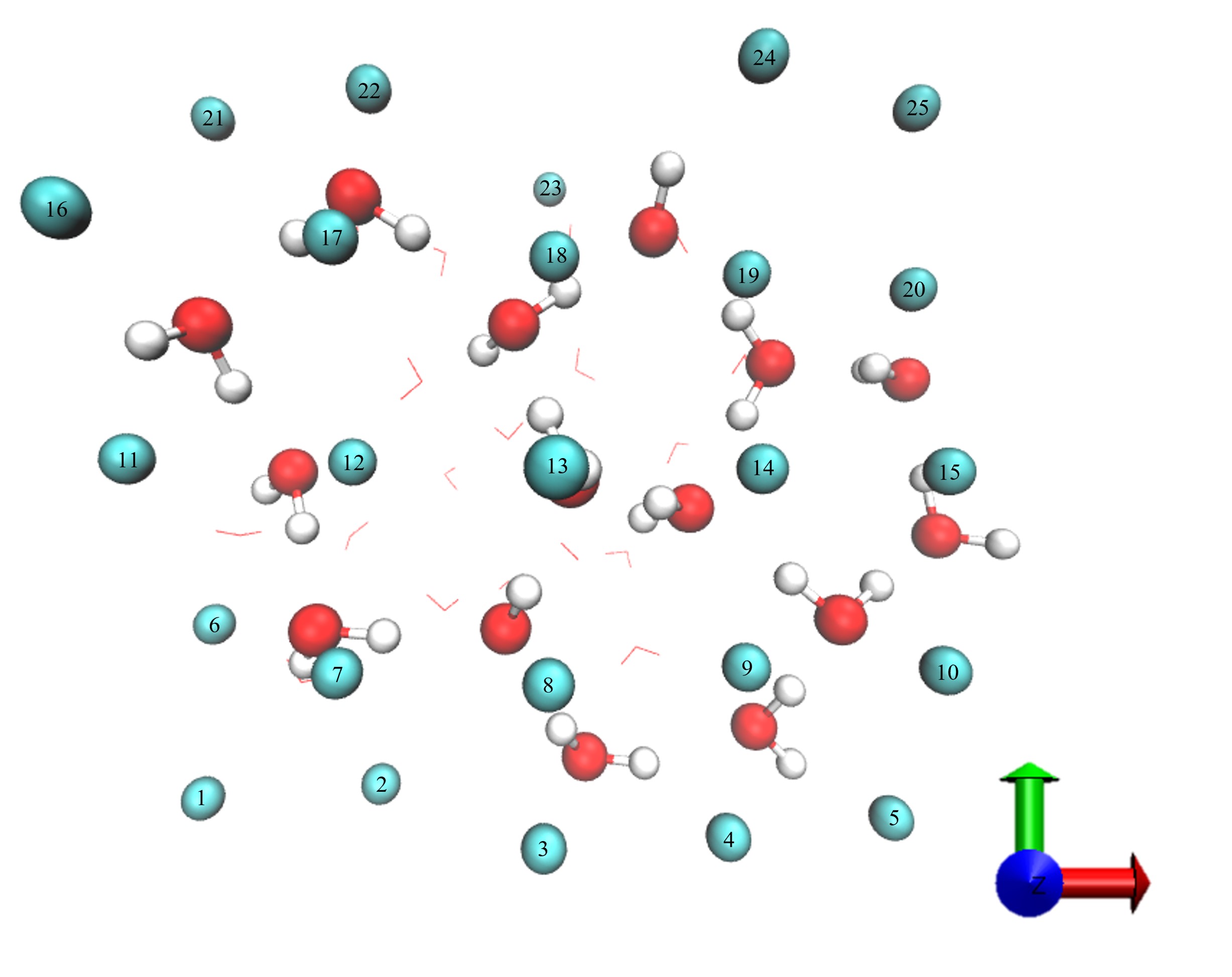}{0.4\textwidth} {(c) sample numbers}}
\caption{Neutral (a) and Charged (b) ASW surface used for the calculations. (c) Blue dots show where the molecules are placed on the ASW surface.}
\label{fig:ASW_slab}
\end{figure}

\begin{figure}[htb!]
\gridline{\leftfig{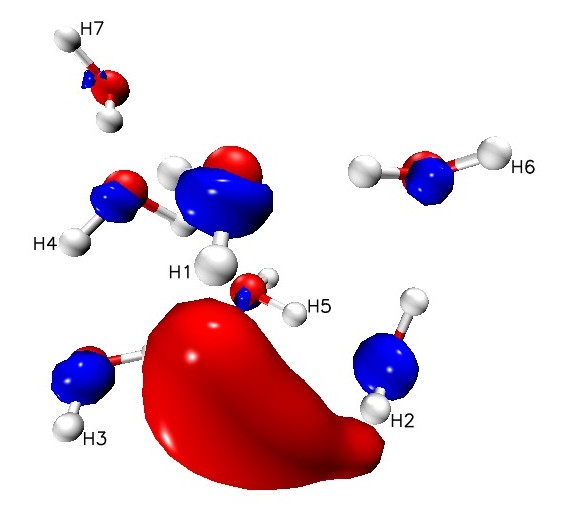}{0.4\textwidth}{(a) HOMO of charged ASW}           
          \rightfig{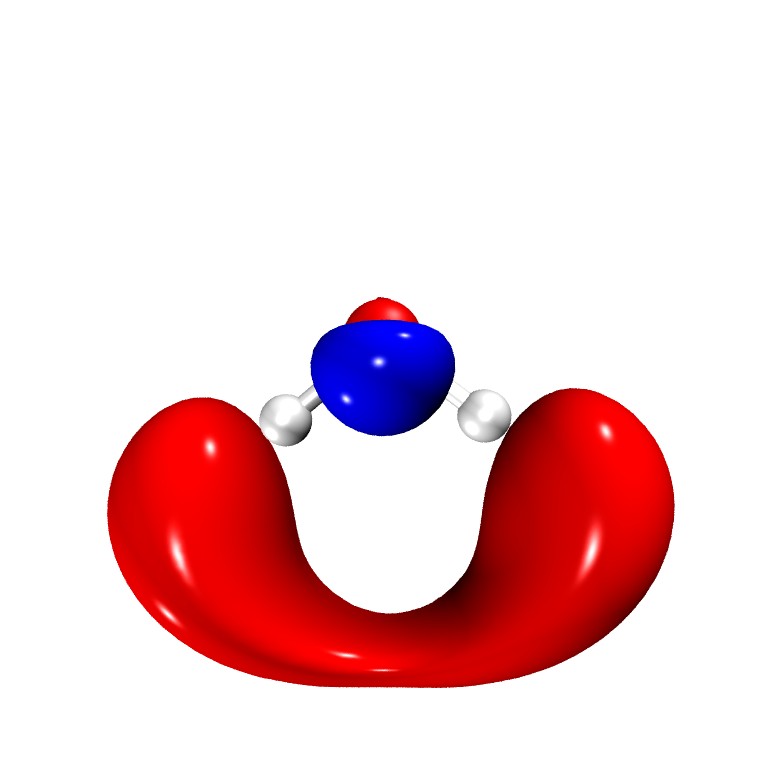}{0.4\textwidth}{(b) LUMO of water}}
\caption{The HOMO of the charged ASW cluster (a), only showing the closest interacting water molecules for clarity. (b) Shows the LUMO of a neutral water molecule.}
\label{fig:ASW_MO}
\end{figure}

Figure \ref{fig:ASW_slab} shows the ASW model used in our calculations. Panels \ref{fig:ASW_slab}a and \ref{fig:ASW_slab}b show the front view of the ASW surface along the y-axis with the z-axis pointing upwards. It can be seen that the optimized charged model has a different configuration compared to the neutral model. Panel \ref{fig:ASW_slab}c shows the sites where we position a molecule on the surface before optimization. Upon further investigation of the charged ASW cluster, we found that the electron is localized in a position where it is stabilized by several H-atoms. Figure \ref{fig:ASW_MO}a shows the molecular orbital which contains the extra electron. This molecular orbital is composed of the overlap from multiple LUMO's of water. Indeed, if we visualize the LUMO of water, as shown in figure \ref{fig:ASW_MO}b, we see that overlapping the lobes on the outside of the molecule will result in a space where an electron can reside. Our analysis of the water molecule's LUMO indicates that it is predominantly made up of hydrogen atomic orbitals. We are now able to determine the atomic orbital structure of the charged ASW HOMO hosting the additional electron, as well as calculate the bond lengths of the dangling hydrogens interacting with this charge. From this analysis, we found that the molecular orbital is mostly composed of the atomic orbitals from the dangling hydrogens. This is analogous to the water LUMO, indicating that it is indeed this molecular orbital where the extra electron resides. Additionally, looking at the bond lengths shown in table \ref{tab:HOMO}, the bond length increases when the dangling hydrogen contribution increases. This is expected since the extra electron is placed in an antibonding orbital of water, decreasing the bond strength, thus increasing its length.

In Table \ref{tab:results} we show the calculated minimum, maximum and average binding energy for each molecule. We opted to show our results in this way since all individual binding sites give rise to their individual particular binding energies due to the amorphous nature of the surface. In the following subsections, the results for each molecule will be presented in more detail. The explanation of these results will then follow in the discussion.

\begin{deluxetable}{c c c c c c c c}[hbt!]
    \tablecaption{Atomic composition of the HOMO of the charged ASW cluster. The O-H bond lengths are also shown, a normal dangling hydrogen bond has an average bond length of 0.9613 \AA.} \label{tab:HOMO}
    \tablehead{\colhead{Atom} & \colhead{Dangling?} & \colhead{Atomic contribution} & \colhead{Bond length (\AA)}} 
    \startdata
        H1 & yes & 53.67\% & 0.9709\\
        H2 & yes & 52.65\% & 0.9703\\
        H3 & yes & 38.85\% & 0.9650\\
        H4 & no & 18.17\% & 0.9726\\
        H5 & no & 17.19\% & 0.9821\\
        H6 & yes & 12.99\% & 0.9618\\
        H7 & yes & 10.87\% & 0.9610\\
        \enddata
\end{deluxetable}

\begin{deluxetable}{c c c c c c c c}[hbt!]
    \tablecaption{Our results for the binding energies (BE) of CO, CH$_4$, and NH$_3$ on a neutral and charged ASW surface. The average is calculated over all 25 binding positions. Units are in K and the energies are without zero point energy correction.} \label{tab:results}
    \tablehead{\colhead{} & \multicolumn3c{Neutral BE} & & \multicolumn3c{Charged BE} \\
    \colhead{Molecule} & \colhead{Min} & \colhead{Max} & \colhead{Average} & & \colhead{Min} & \colhead{Max} & \colhead{Average}} 
    \startdata
        CO & 1205 & 2144 & 1621 & & 1277 & 5734 & 2161 \\
        CH$_4$ & 920 & 1771 & 1312 & & 1071 & 1564 & 1378 \\
        NH$_3$ & 4308 & 14768 & 7310 & & 4479 & 7769 & 5926 \\
        \enddata
\end{deluxetable}

\subsection{CO} \label{subsec:CO}

\begin{figure}[htb!]
\gridline{\fig{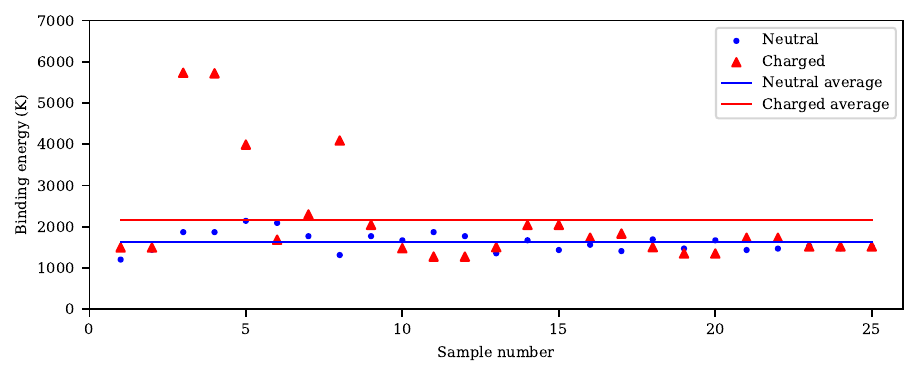}{\textwidth}{(a) CO Binding energy results}}
\gridline{\leftfig{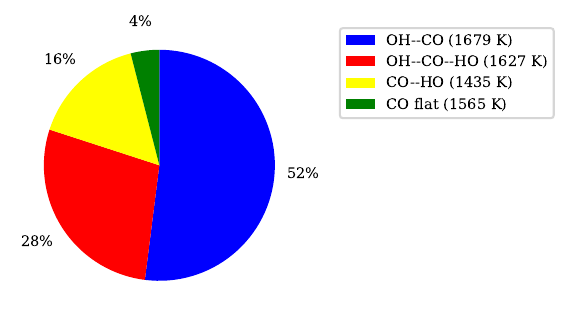}{0.49\textwidth}{(b) Neutral end configurations}           
          \rightfig{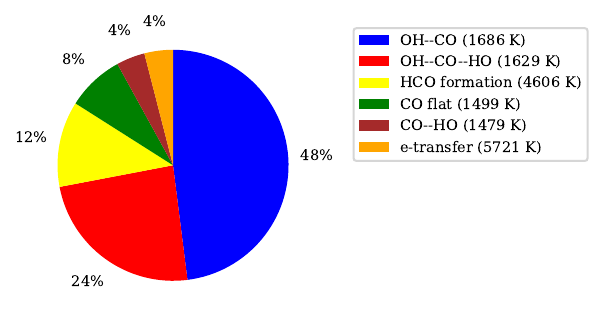}{0.49\textwidth}{(c) Charged end configurations}}
\caption{Results from the CO calculations. (a) shows the neutral and charged binding energies, without zero point energy correction. (b) and (c) show the distributions over the different end configurations.}
\label{fig:CO_results}
\end{figure}

Figure \ref{fig:CO_results} shows the results for CO adsorption. Panel \ref{fig:CO_results}a shows the binding energies for the neutral and charged model. Panels \ref{fig:CO_results}b and \ref{fig:CO_results}c show the distribution of the results in the different end configurations observed. In panel \ref{fig:CO_results}a every data point represents one of the 25 calculations, thus showing all results. According to graph \ref{fig:CO_results}a, the neutral binding energies are found to lie in the range 1205 K - 2144 K. The charged binding energies are found in a wider range, viz. between 1277 K and 5734 K.

\begin{figure}[htb!]
\gridline{\leftfig{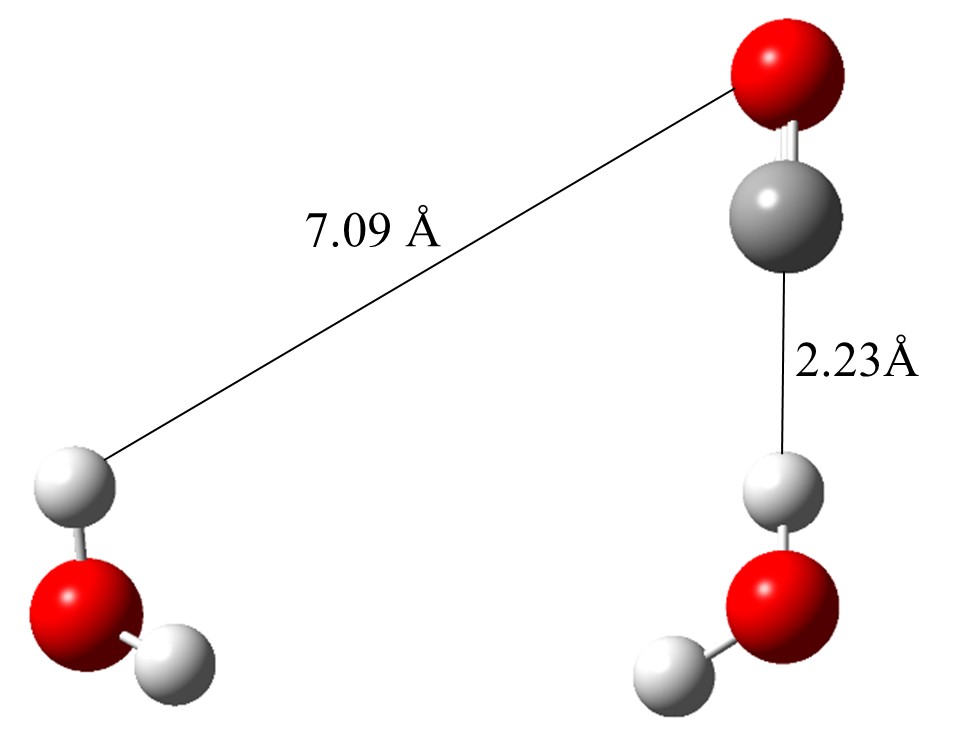}{0.4\textwidth}{(a) End configuration OH--CO}           
          \rightfig{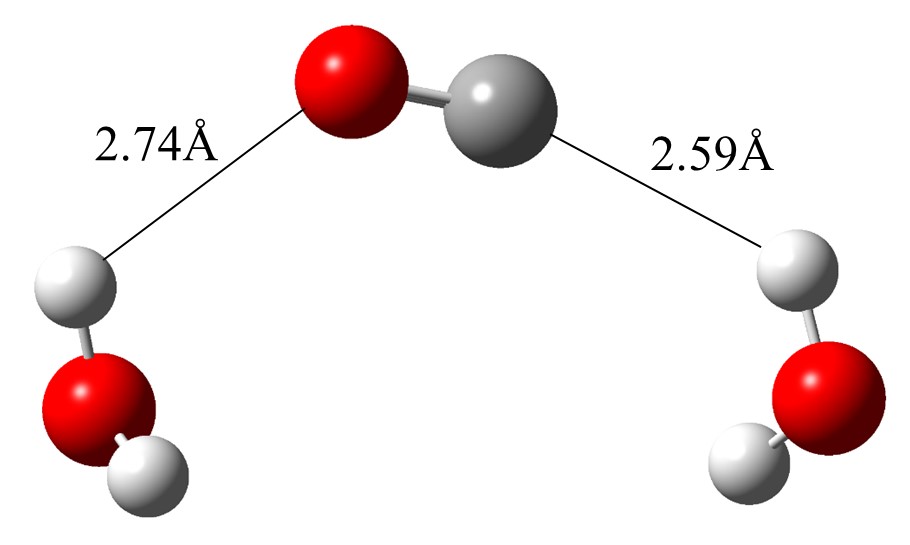}{0.4\textwidth}{(b) End configuration OH--CO--HO}}
\caption{Interaction between CO and the ASW surface for end configurations OH--CO and OH--O-C--HO.}
\label{fig:CO_distances}
\end{figure}

The distribution of end configurations on the neutral surface (\ref{fig:CO_results}b) shows that there are four ways for CO to bind to the neutral ASW surface. The first configuration (denoted OH--CO) consists of only one interaction with the surface, viz. the carbon atom from CO interacts with a dangling H from the surface. This configuration yields an average binding energy of 1563 K. In the second configuration (denoted OH--CO--HO), the CO interacts with two separate dangling H's. Here the C and O atoms each interact with a separate dangling H. Figure \ref{fig:CO_distances} shows the interaction between the CO molecule and the ASW surface for OH--CO and OH--CO--HO. The difference between the two configurations is the interaction length between the O-atom of CO and a second dangling H. The OH--CO--HO configuration shows a larger binding energy of 1616 K due to this additional interaction. The third configuration (denoted CO Flat) shows the CO molecule lying flat on the ASW surface, without any clear interaction with a dangling H. Instead, the CO molecule is close enough to the surface to interact with multiple water molecules with weaker individual interactions. These weaker interactions explain the lower binding energy of 1500 K. The fourth (CO in cavity) configuration is observed when the CO molecule binds in a cavity. In this configuration, there are no dangling H's in the proximity of the CO molecule. This configuration shows the lowest binding energy (1437 K) because it only interacts with water molecules which themselves are H-bonded to neighboring water molecules.

\begin{figure}[htb!]
\gridline{\leftfig{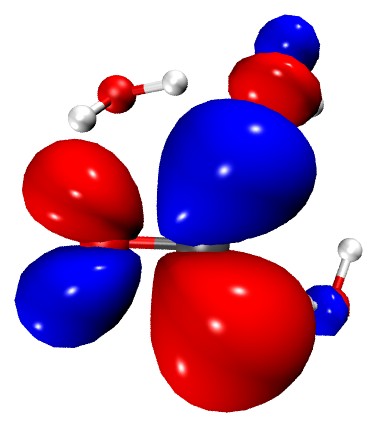}{0.3\textwidth}{(a) e-transfer 
          HOMO}           
          \rightfig{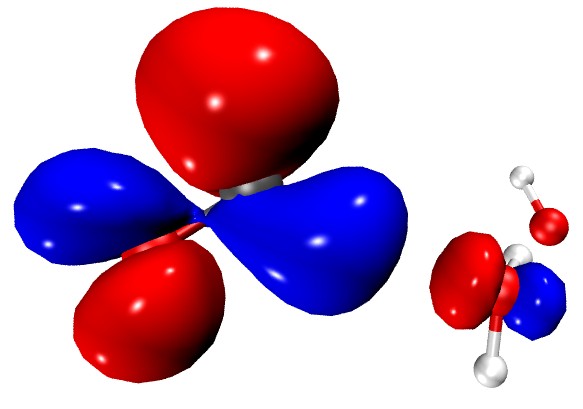}{0.4\textwidth}{(b) HCO formation HOMO}}
\caption{HOMO's of the e-transfer and HCO formation end configurations, showing the clear transfer of the extra electron from the surface to the CO molecule.}
\label{fig:CO_HOMO}
\end{figure}

The end configurations for the charged surface are divided in the same way as the neutral end configurations. Additionally, there are two more possible configurations (denoted as HCO formation and e-transfer). Comparing the four recurring end configurations in figure \ref{fig:CO_results}, we see that the binding energies do not change significantly when a charge is added to the surface. There are, however, two new end configurations corresponding to the four higher binding energies shown in figure \ref{fig:CO_results}a. When the CO approaches the charged surface, an electron transfer can occur from the ASW surface towards the CO. Additionally, negatively charged CO can spontaneously abstract a H-atom from the surface, thus forming HCO. Both the e-transfer and HCO formation show a significant increase in binding energy. To determine the binding energy after HCO formation, we calculate it as the binding energy of HCO on the ASW surface with one hydrogen atom removed. This method retains the concept of determining the binding energy of a molecule on a surface. We also visualized the HOMO of these two configurations, shown in figure \ref{fig:CO_HOMO} to verify if the electron has indeed transferred to the CO molecule. In both cases we clearly see the extra electron residing in the antibonding orbital of CO or HCO.

\begin{figure}[htb!]
\gridline{\leftfig{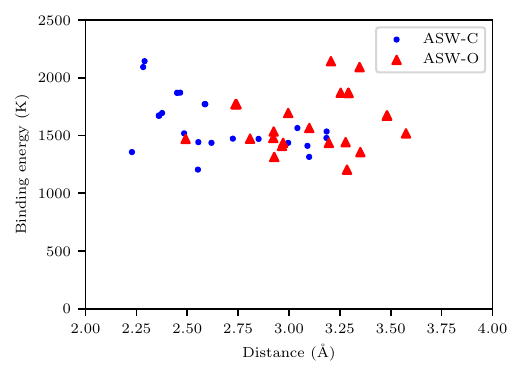}{0.49\textwidth}{(a) Minimum ASW-CO distance on the neutral surface}           
          \rightfig{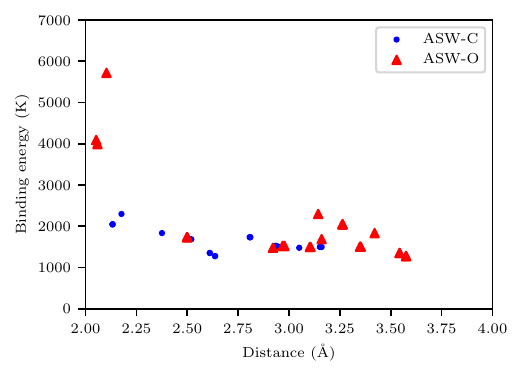}{0.49\textwidth}{(b) Minimum ASW-CO distance on the charged surface}}
\gridline{\leftfig{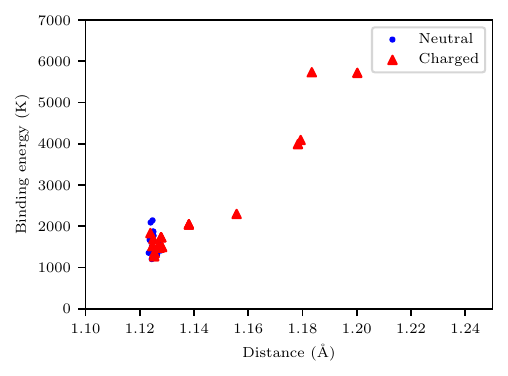}{0.49\textwidth}{(c) Bond length of CO}}
\caption{Intermolecular distances between CO and the ASW surface (a and b) and the bond length of CO after adsorption (c).}
\label{fig:ASW-CO}
\end{figure}

The intermolecular and intramolecular distances can be used as an indicator of the interaction strength. In figure \ref{fig:ASW-CO} we show the binding energies of all configurations in relation to multiple distances. Examining figure \ref{fig:ASW-CO}a, it is evident that on a neutral surface, an arrangement where the C-atom is closer to the surface and the O-atom is farther away results in higher binding energies. This indicates that the OH--CO configuration depicted in figure \ref{fig:CO_results}b yields the optimal interaction for CO, this is indeed the configuration with the highest average binding energy. In the case of a charged surface (\ref{fig:ASW-CO}b), a similar trend is observed except at the highest binding energies, where small distances between ASW and the O atom are observed. These configurations involve electron transfer and HCO formation, during which an H atom is extracted from the surface and subsequently positions itself near the CO molecule. Finally, it is observed that the bond length of CO increases as the binding energy rises (figure \ref{fig:ASW-CO}c). This happens when CO forms a strong bond with the charged surface, it attracts some electron density into its anti-bonding orbital, leading to the lengthening of the CO bond.

\pagebreak
\subsection{CH$_4$} \label{subsec:CH4}

\begin{figure}[htb!]
\gridline{\fig{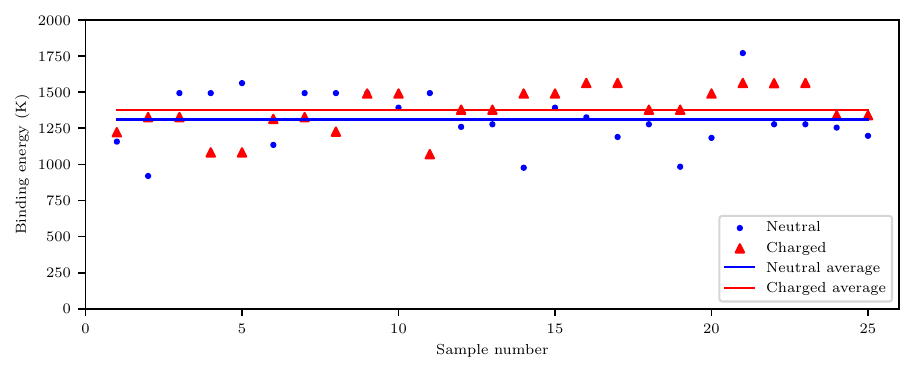}{\textwidth}{(a) CH$_4$ Binding energy results}}
\gridline{\leftfig{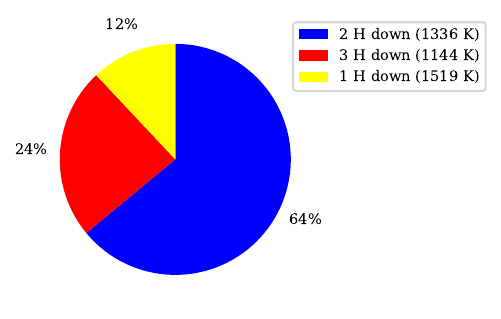}{0.49\textwidth}{(b) Neutral end configurations}           
          \rightfig{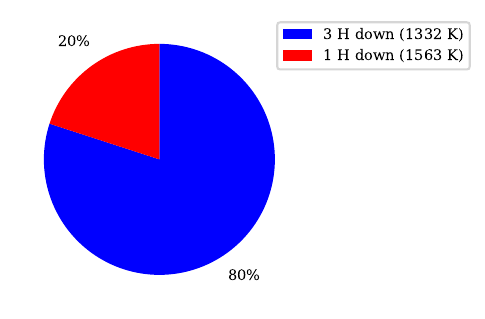}{0.49\textwidth}{(c) Charged end configurations}}
\caption{Results from the CH$_4$ calculations. (a) shows the neutral and charged binding energies, without zero point energy correction. (b) and (c) show the distributions over the different end configurations.}
\label{fig:CH4_results}
\end{figure}

Figure \ref{fig:CH4_results} shows the results of the calculations for CH$_4$. Comparing the binding energies in Table \ref{tab:results}, we see that CH$_4$ has the lowest binding energies. CH$_4$ interacts with the surface through dispersion forces. On the charged ASW surface, the binding energies of CH$_4$ on a charged ASW surface are within a narrow spread between 1071 K and 1564 K. On the neutral surface, the spread in binding energies is between 753 K and 1972 K. The pie chart in panels \ref{fig:CH4_results}b and \ref{fig:CH4_results}c presents the distribution in the different end configurations. The end configurations are divided based on the number of H-atoms from CH$_4$ which are closer to the surface than the C-atom. For both the neutral and charged systems, we see an increase in the binding energy as the number of H-atoms close to the surface decreases. Additionally, when the surface is charged, the CH$_4$ molecule prefers to interact with the surface with three H-atoms with 80\% of the end configurations in this position.

\begin{figure}[htb!]
\fig{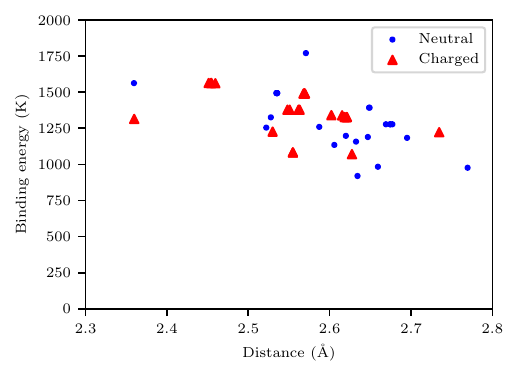}{0.49\textwidth}{}
\caption{Intermolecular distance between the ASW surface and CH$_4$ for every neutral and charged configuration.}
\label{fig:ASW-CH4}
\end{figure}

Figure \ref{fig:ASW-CH4} shows the smallest intermolecular distance between the ASW surface and the CH$_4$ molecule. This is a good indicator for how close to the surface the molecule is positioned in the end configuration. We see that for higher binding energies, the molecule will be closer to the surface. This is reasonable, as the strength of dispersion forces increases when the distance between two species shortens.

\subsection{NH$_3$} \label{subsec:NH3}

\begin{figure}[htb!]
\gridline{\fig{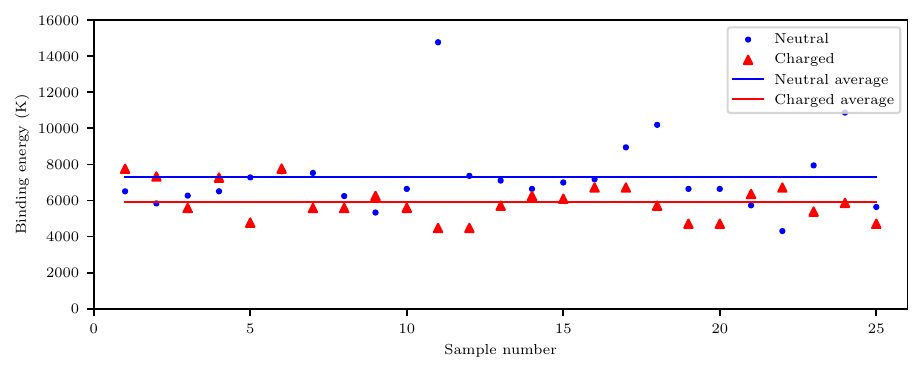}{\textwidth}{(a) NH$_3$ Binding energy results}}
\gridline{\leftfig{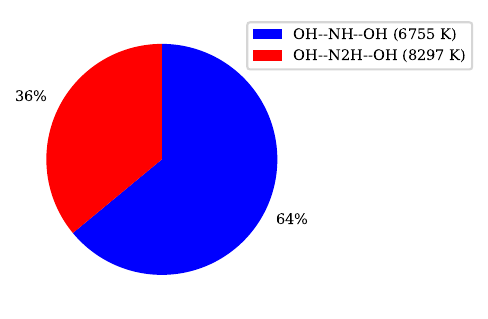}{0.49\textwidth}{(b) Neutral end configurations}           
          \rightfig{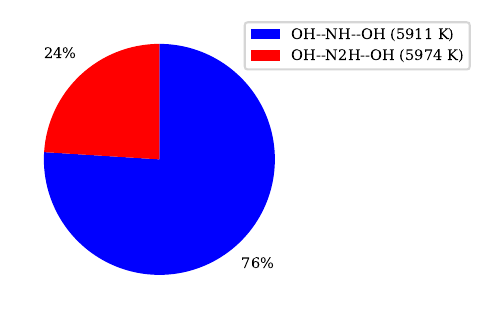}{0.49\textwidth}{(c) Charged end configurations}}
\caption{Results from the NH$_3$ calculations. (a) shows the neutral and charged binding energies, without zero point energy correction. (b) and (c) show the distributions over the different end configurations.}
\label{fig:NH3_results}
\end{figure}

The binding energies of NH$_3$ strongly depend on the number of hydrogen bonds formed with the surface as NH$_3$ is both a hydrogen bond donor and acceptor. Table \ref{tab:results} shows that NH$_3$ has the highest neutral binding energy, in the range of 4308 K to 14768 K. The binding energies on the charged ASW surface have a smaller spread and are in general also weaker, showing a range between 4479 and 7769 K. Figure \ref{fig:NH3_results}a shows the results for the neutral and charged surfaces.

As mentioned, NH$_3$ can act both as a hydrogen bond donor and an acceptor. Thus, there are two important interactions between NH$_3$ and the surface. The first is a hydrogen bond between the N atom and a dangling H from the surface. We refer to this interaction as OH--N. The second is a hydrogen bond between an O atom from the surface and a H atom from NH$_3$, which we denote as H--OH. Figure \ref{fig:NH3_results}b and \ref{fig:NH3_results}c show the distribution in these configurations. For the neutral surface, the first configuration (OH--NH--OH) has an average binding energy of 6755 K and has one OH--N and one H--OH interaction. The second configuration (OH--N2H--OH) has one OH--N interaction and two H--OH interactions. This yields the highest average binding energy of 8297 K.

The end configurations on the charged surface are also divided according to the key interactions discussed earlier. Figure \ref{fig:NH3_results}c shows the distribution and average binding energies. The same observations as for the neutral surface can be made for the charged configurations. We see that the binding energy increases when more interactions occur between the NH$_3$ molecule and the charged surface. However, the average charged binding energy per end configuration is lower than for the neutral counterpart. Furthermore, from figure \ref{fig:NH3_results}a it can be observed that the average binding energy on the charged surface is lower than on the neutral surface.

\begin{figure}[htb!]
\gridline{\leftfig{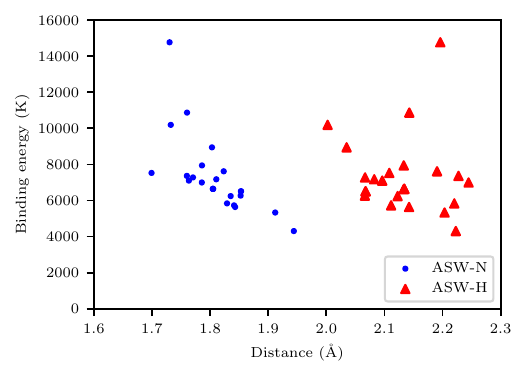}{0.49\textwidth}{(a) Minimum ASW-NH$_3$ distance on the neutral surface}           
          \rightfig{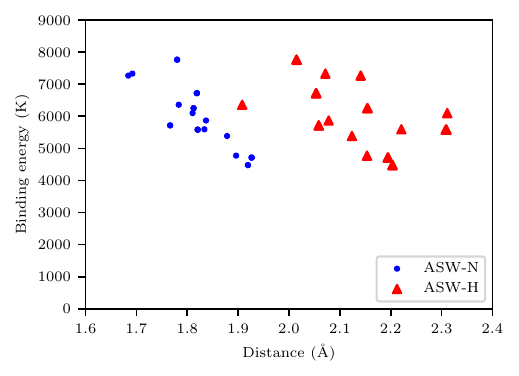}{0.49\textwidth}{(b) Minimum ASW-NH$_3$ distance on the charged surface}}
\caption{Intermolecular distances for the adsorption of NH$_3$ on an neutral (a) and a charge (b) ASW surface .}
\label{fig:ASW-NH3}
\end{figure}

Figure \ref{fig:ASW-NH3} shows the intermolecular distance between the ASW surface and the NH$_3$ molecule. It can be seen that for both a neutral and charged surface the binding energy increases when the intermolecular distance decreases. This shows that when the molecule is closer to the surface both its H-bond donating and accepting properties are strengthened, which increases the binding energy. This is also reflected in the configurations in figures \ref{fig:NH3_results}b and \ref{fig:NH3_results}c. When NH$_3$ is close to the surface, more H-bonds are possible which increase the binding energy. Moreover, we note that a charged surface results in a longer average hydrogen bond length of 1.8268 \AA\ compared to the 1.8084 \AA\ observed on a neutral surface. The individual atomic charges related to the hydrogen bond connecting the N-atom to the ASW surface also change. When nitrogen is adsorbed on a neutral surface, its average charge is -0.92, which is more negative than the average charge of -0.86 on the neutral surface. Additionally, the charge on the dangling hydrogen on the surface undergoes a minor change, decreasing from an average of 0.46 on a neutral surface to 0.44 on a charged surface. Consequently, the difference in charge, which influences the strength of the hydrogen bond, decreases from an average value of 1.38 to 1.30 when the surface is charged. This indicates a weakening of the H-bonds, which is indeed what we observe from the binding energy results (table \ref{tab:results}).

\subsection{Zero point energy} \label{subsec:ZPE}

\begin{figure}[htb!]
\fig{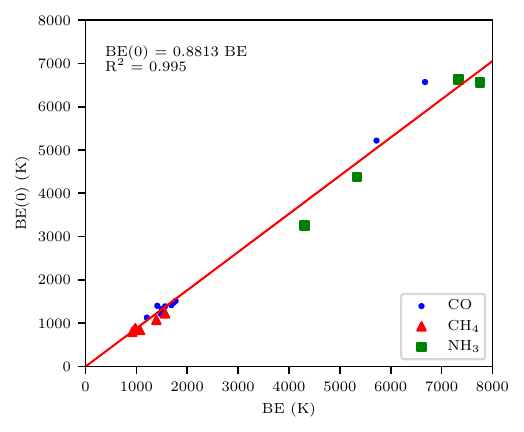}{0.5\textwidth}{}
\caption{Linear regression between the binding energy (BE) and the zero point energy corrected binding energy (BE(0)) for CO, CH$_4$ and NH$_3$.}
\label{fig:ZPE}
\end{figure}

For each distinct final configuration, we calculated the zero point energy correction related to the binding energies of CO, CH$_4$, and NH$_3$. Using these frequency calculations we were able to confirm every end configuration as a true minima of the potential energy surface. Figure \ref{fig:ZPE} shows the relation between the binding energy (BE) and the zero point energy corrected binding energy (BE(0)). From this plot we are able to extrapolate the zero point energy correction, so that BE(0) can be estimated using the correlation $BE(0) = 0.8813$ $BE$. This study aims to explore the relative differences in binding energies between neutral and charged surfaces. Therefore, we will focus on uncorrected binding energies in the discussion, as we chose not to calculate the zero point energy correction for each sample to manage computational costs.

\section{Discussion} \label{sec:discussion}
\subsection{Describing charged systems in DFT} \label{subsec:DFT}

Before discussing the results, we first discuss some important aspects of using DFT to describe a charged system. Exchange-correlation functionals in DFT are affected by many-electron self-interaction error \citep{perdew1981}, derivative discontinuity \citep{perdew1982}, and delocalization error \citep{autschbach2014}. These errors may result in the instability of anions \citep{vydrov2007}, the description of anions with only a fraction of the electron bound \citep{jensen2010} and a too high Highest Occupied Molecular Orbital (HOMO) \citep{amati2019}. When employing DFT for the analysis of charged systems, these limitations must be considered. We therefore first recalculate the binding energies of the HCO$^+$ cation, investigated previously by Rimola et al., using our method \citep{Rimola2021}. Rimola et al. used clusters of different sizes and charge localizations. The largest cluster they used consisted of 24 water molecules, which is closest to our cluster of 33 water molecules. Our cluster localizes the charge, so we will compare our results with those of the 24 water molecules cluster with a localized charge. This cluster yielded a binding energy of 7.98 eV. Using our method, we found an average binding energy of 7.87 eV. Comparing our value (7.87 eV) with the value in the literature (7.98 eV), we find a good correspondence with only a difference of 1. 40\%. Since there is good agreement between our charged results and those from Rimola et al. we can confirm that the charged binding energies are calculated correctly. This is especially true for the CO calculations, since the HCO$^+$ cation is a derivative of the CO molecule and thus has a similar electronic structure. Additionally, the CO molecule has an electronic structure which is most difficult to describe of the three molecules reported in this paper. Therefore, we can assume that the other two molecules can also be calculated with good accuracy using our method.

\subsection{Neutral binding energies} \label{subsec:Neutral}

\begin{figure}[htb!]
\gridline{\leftfig{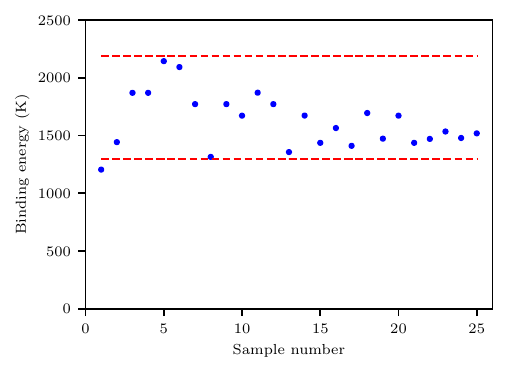}{0.49\textwidth}{(a) CO}           
          \rightfig{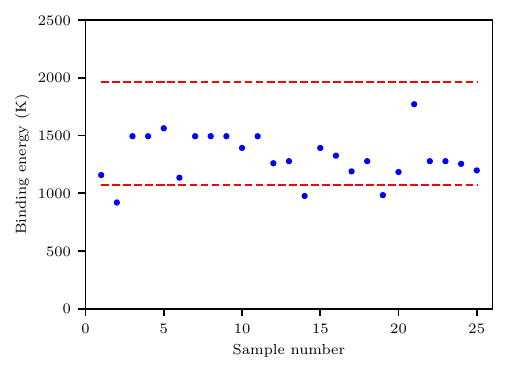}{0.49\textwidth}{(b) CH$_4$}}
\gridline{\leftfig{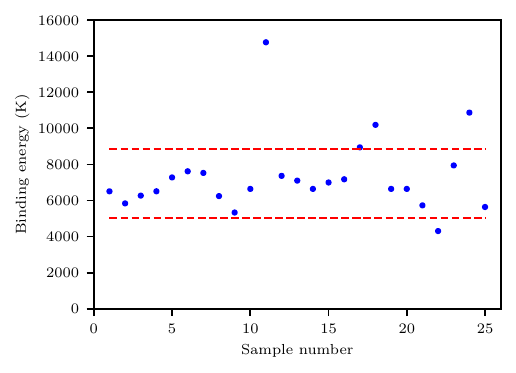}{0.49\textwidth}{(c) NH$_3$}}
\caption{Comparison of our computed neutral binding energies on the ASW surface with the results from Ferrero et al. (red dotted lines). All energies do not include zero point energy correction.}
\label{fig:Comparison}
\end{figure}

Every binding site on the ASW slab used for our calculations is unique, which is characteristic of an amorphous surface. Consequently, the binding energies of the three molecules lie within a wide range as reported in Table \ref{tab:results}. This wide spread in binding energies is already well-known in the literature \citep{bovolenta2022}. It does, however, mean that comparing our results to those in the literature is a delicate matter. The calculated binding energies are - within a given set of computational approximations - not only determined by the used ASW surface but also by the number of binding sites accounted for and thus the number of calculated binding energies. Taking a large sample size increases the possibility of finding binding opportunities that are not yet described in the literature, making a direct comparison of the average binding energies difficult.

We compare our results for the neutral binding energies to the binding energies calculated by Ferrero et al. It is important to note that we simulated 25 configurations per molecule, while Ferrero et al. calculated five cases for CO and CH$_4$ and seven cases for NH$_3$. Figure \ref{fig:Comparison}a shows our calculated results for CO. Each of the dots represents a particular calculation result of ours, while the red dotted lines are the minimum and maximum values from Ferrero et al. They reported that the CO binding energy is between 1299 K and 2189 K. Overall, we see a very good agreement between our results and those from Ferrero et al.

As shown in figure \ref{fig:Comparison}b, we observe the same trend for CH$_4$ as for CO. We find that our binding energies mostly lie between the minimum and maximum reported values from Ferrero et al.

Lastly, in figure \ref{fig:Comparison}c, we compare our results for NH$_3$ with the values reported by Ferrero et al. Here, we again have a similar trend where most energies lie between the minimum and maximum reported values, but there are three cases where the binding energy is significantly higher than the maximum reported by Fererro et al. At these higher binding energies, the NH$_3$ molecule maintains the same interaction with the surface. Yet the closest water molecules align such that the NH$_3$ molecule can engage more strongly with the dangling H. This is confirmed by the intermolecular distance between the N-atom and the dangling H, which is shorter than for the other configurations.

\subsection{Charged binding energies}

\begin{figure}[htb!]
\gridline{\leftfig{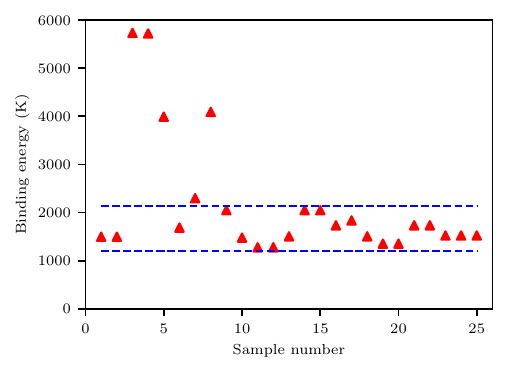}{0.49\textwidth}{(a) CO}           
          \rightfig{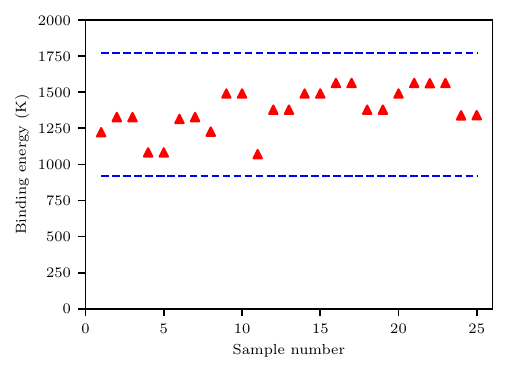}{0.49\textwidth}{(b) CH$_4$}}
\gridline{\leftfig{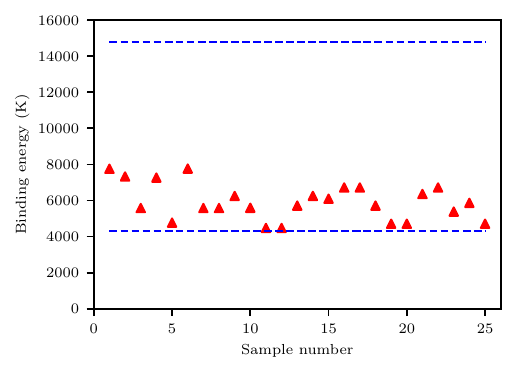}{0.49\textwidth}{(c) NH$_3$}}
\caption{Comparison of our computed charged binding energies on the ASW surface with our neutral results (blue dotted lines). All energies do not include zero point energy correction.}
\label{fig:Comparison_charged}
\end{figure}

\begin{figure}[htb!]
\gridline{\leftfig{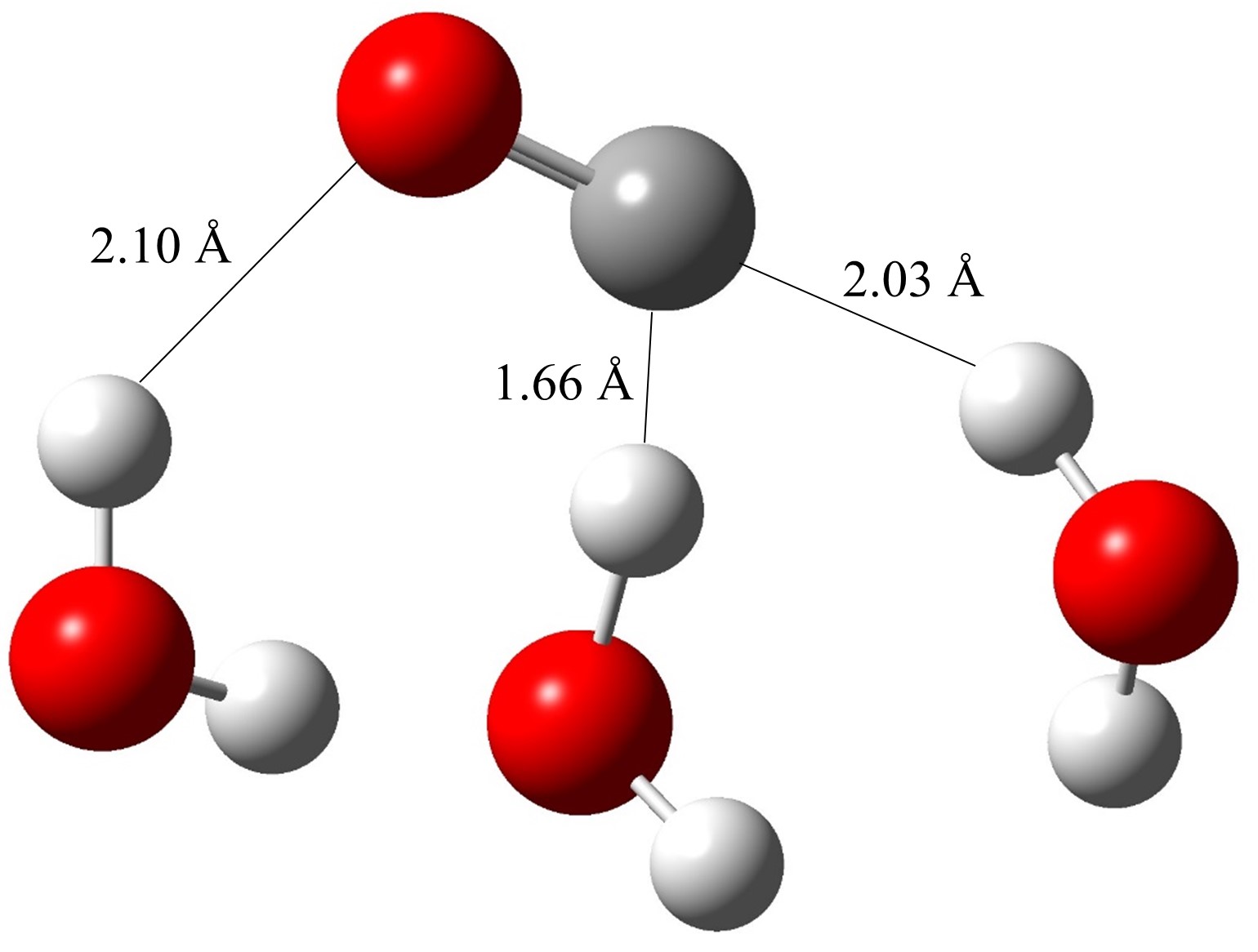}{0.40\textwidth}{(a)             e-transfer}           
          \rightfig{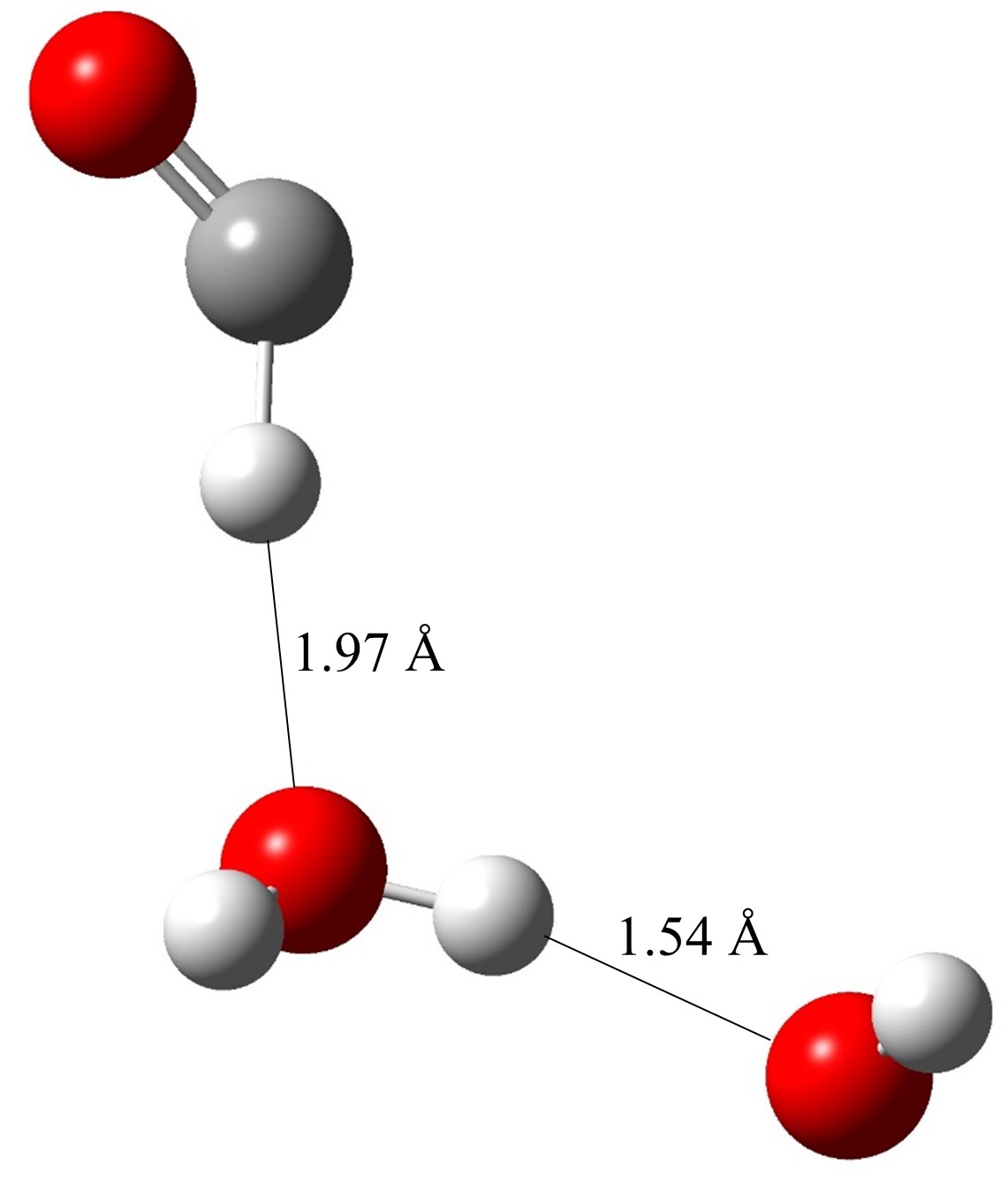}{0.35\textwidth}{(b) HCO formation}}
\caption{The two possible end configurations when CO interacts with the extra electron from the surface. (a) Shows an electron transfer from the surface towards the CO molecule and (b) shows a spontaneous H-abstraction forming HCO. Only the closest water molecules form the surface are shown for simplicity.}
\label{fig:CO_interaction}
\end{figure}

Figure \ref{fig:Comparison_charged} compares our results for the charged system to those for the neutral system, in order to identify significant changes in binding energy. For CO we see that most of the binding energies for the charged system are within the boundaries of the energies for the neutral system. Figure \ref{fig:CO_results}c indicates that the charged binding energies can be split into six groups, based on their interaction with the surface. The higher binding energies of CO can be attributed to the cases in which CO interacts directly with the charge on the surface. In total, four samples have an elevated binding energy. One of these processes is due to an electron transfer of the extra electron from the surface towards the CO. In the other three cases, spontaneous H-abstraction occurs, forming HCO. Based on the computed multiplicities, the additional electron remains on the HCO molecule, essentially creating the HCO radical. Figure \ref{fig:CO_interaction} shows the two end configurations associated with the interaction between CO and the extra electron from the surface. \ref{fig:CO_interaction}a shows the end configuration after electron transfer. It is clear that the water molecules retain all of their H-atoms and CO holds the extra electron. The surrounding dangling hydrogens help to stabilize the extra electron on CO. For the second configuration, panel \ref{fig:CO_interaction}b shows the formation of HCO after H-abstraction from the surface. Here, the HCO is a radical and interacts with the surface with its H-atom. Additionally, we see that after H-abstraction, proton transfer occurs between the water molecules to stabilize the $^-$OH moiety created in the surface.

Figure \ref{fig:Comparison_charged}b shows the results for adsorption of CH$_4$ on a charged ASW surface. We see that all binding energies (orange dots) fall in between the blue dotted lines, which are the minimum and maximum calculated energies for the neutral surface. The average charged binding energy is 1378 K, which differs only slightly from the average neutral binding energy of 1312 K (i.e., a 5\% increase). Upon further inspection of the final configurations, we do not observe any interaction between CH$_4$ and the charge on the surface. However, we do see a change in the distribution of end configurations after the surface is charged, but without any significant change in binding energy. We can therefore conclude that the binding energy of CH$_4$ on an ASW surface is not influenced by surface charge.

Finally, we see in figure \ref{fig:Comparison_charged}c that the binding energies of NH$_3$ on the charged surface are near the lower boundary of the neutral calculations. From figure \ref{fig:NH3_results}a we indeed see that the average charged binding energy is 5926 K, while the average neutral binding energy is 7310 K (i.e., a 19\% decrease). As mentioned earlier, the binding energy of NH$_3$ primarily relies on the hydrogen bonds that are formed. By examining the various final configurations (figures \ref{fig:NH3_results}b and \ref{fig:NH3_results}c), the average binding energy decreases in all cases. This reduction is attributed to the weakening of hydrogen bonds. By charging the surface, we introduce a repulsive force between the negatively charged surface and the nitrogen atom in the NH$_3$ molecule.

\subsection{Astrophysical implications}

To conclude the discussion we want to point to the astrophysical implications of our results. First, we know that icy dust grains in the ISM can be negatively charged \citep{draine1987}. There are a few mechanisms for this charging, of which the most important is the collision of dust grains with thermal electrons and ions from the gas phase \citep{ivlev2015}. This process is sometimes referred to as cold plasma charging. In the literature, it is already mentioned that the charge on interstellar grains have important consequences for the chemical and dynamical evolution of molecular clouds \citep{ivlev2015}. The charge mainly affects processes like dust coagulation \citep{okuzumi2009}, grain-catalyzed reactions \citep{mestel1956}, and the amount of gas-phase depletion \citep{spitzer1941}.

The above-mentioned processes are often studied using grain surface modeling techniques. These models include gas-surface interactions to describe the evolution of molecular abundances over time. The different processes described in these models are accretion, desorption, reaction, diffusion, bulk processes and photoprocesses \citep{cuppen2017}. Two of these processes are heavily influenced by the binding energy of species on the ASW surface. The first process is thermal desorption. The desorption energy in these models is typically taken as $E_{des} = -E_{bind}$. The desorption rate of species $x$ at surface temperature $T_s$ is typically given

\begin{equation} \label{eq:desorption}
    k_{des, x} = \nu_{trial} exp(\frac{-E_{bind, x}}{k_b T_s})
\end{equation}

The second process is the diffusion of species over the surface. There is again a linear relation between the diffusion energy and the binding energy, $E_{diff, x} = \alpha E_{bind, x}$ with $\alpha < 1$. There is an analogous equation for the rate of diffusion of species $x$ on a surface with temperature $T_s$, shown in equation \ref{eq:diffusion}.

\begin{equation} \label{eq:diffusion}
    k_{diff, x} = \nu_{trial} exp(\frac{-E_{diff, x}}{k_b T_s})
\end{equation}

Both rates depend exponentially on the binding energy of a species. This means that when the binding energies of species change due to the surface being charged, the rates of these processes will dramatically change as well. This change results in different molecular abundances as calculated by the model.

In a study by Penteado et al. the sensitivity of grain surface chemistry to the binding energies on ASW surfaces was analyzed \citep{penteado2017}. They calculated species abundances using standard astrochemical modeling techniques, implementing binding energies and their uncertainties. The uncertainty on a binding energy is either a known value from previous work or a fixed value, e.g. 500 K for binding energies above 1000 K and half the binding energy if it is below 1000K \citep{penteado2017}. With these uncertainties, these authors found a large influence on the abundance of species. Comparing this with our results, we can assume that these will likely lead to significant changes in the abundance of species.

\begin{figure}[htb!]
\fig{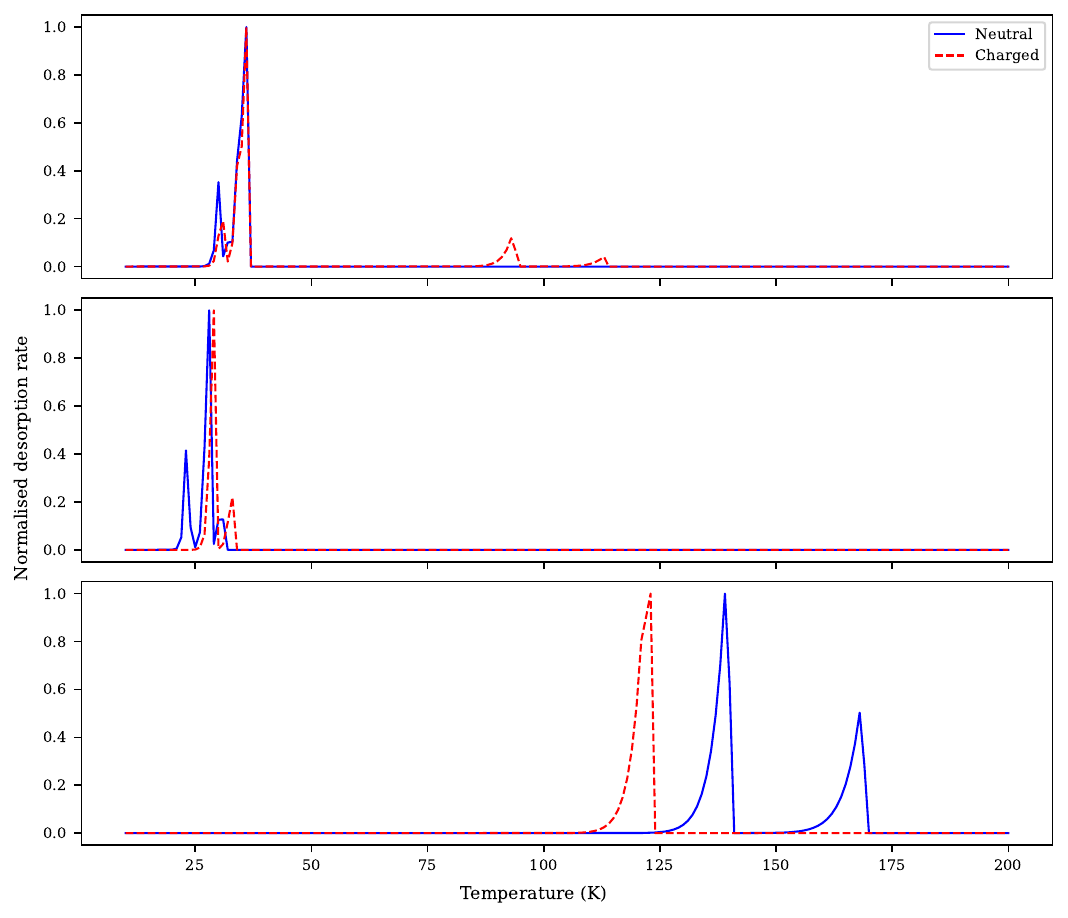}{\textwidth}{}
\caption{Normalised desorption rates for CO (a), CH$_4$ (b) and NH$_3$ (c) when heating an ASW surface from 10 K to 400 K in 10$^5$ years. The blue lines show the desorption rate on a neutral surface, the orange dotted lines show the desorption rate on charged surfaces.}
\label{fig:Model}
\end{figure}

To provide an example of what the impact could be when implementing charged binding energies, we employed the toy model outlined in Ferrero et al.'s paper and made slight modifications to fit our study \citep{Ferrero2020}. We assume a monolayer of CO, CH$_4$ or NH$_3$ without any lateral interactions between the molecules. We then heat up this layer starting form 10 K untill it reaches 400 K in a timespan of 10$^5$ years. This heating pattern is closely related to the heating of a collapsing Solar-like protostar. We can now calculate the desorption rates (using equation \ref{eq:desorption}) at the different temperatures. In figure \ref{fig:Model} the results are shown from these simple calculations. At the start the binding energies of the molecules in the monolayer are distributed as can be seen in figures \ref{fig:CO_results}, \ref{fig:CH4_results} and \ref{fig:NH3_results} for CO, CH$_4$ and NH$_3$ respectively. The desorption rates are also normalised to simplify the comparisons. In the graphs, multiple peaks are seen which are a result of the different binding energies used in the calculation. 

For CO, we see a slight decrease in height for the first peak at 30 K and for the charged surface, a new peak at 110 K appears. This shows that more CO will desorb at lower temperatures on a neutral surface while on a charged surface, some of the CO will stay on the ASW surface until much higher temperatures are reached. 

CH$_4$ does not show much variation in desorption rates between a neutral and charged surface, as was expected from our results. We see a large peak, indicating a lot of desorption at 30 K for both neutral and charged surfaces. Additionally, we see a slight variation in the smaller peaks, where the neutral surface desorbs CH$_4$ at 25 K, while CH$_4$ on a charged surface desorbs at 35 K. Concluding, CH$_4$ that binds to neutral surfaces tends to desorb at somewhat lower temperatures compared to when it is adsorbed on charged surfaces. 

NH$_3$ shows quite a different behavior. On a neutral surface it will desorb at higher temperatures than on a charged surface. For a neutral surface we see two peaks, the largest at 140 K and a smaller at 170 K. On a charged surface, only one large peak is found at 125 K. These simplified findings merely suggest the potential effects of incorporating charges into astrochemical models and already show varied desorption behaviors among the three molecules examined in our study.

Modern gas-grain models often employ single binding energies to describe the behavior of molecules on dust grains. Previous studies have already shown that a binding energy distribution is more accurate in describing grain processes such as desorption and diffusion \citep{Ferrero2020,Tinacci2022}. The implementation of a fraction of charged binding energies in these distributions would bring the model even closer to reality. This improvement enables us to more accurately represent gas abundances in warmer conditions and, for instance, predict the snow lines of protoplanetary disks with greater accuracy.

\section{Conclusion} \label{sec:conclusion}

In this work, we present the calculated binding energies of three relevant interstellar molecules (CO, NH$_3$ and CH$_4$) on a neutral and a charged ASW surface. While the binding energies of small interstellar molecules on neutral surfaces have previously been investigated, the study of these molecules on charged ASW surfaces has not yet been conducted. Because molecular clouds are weakly ionized plasmas, they contain not only neutral species, but also charged species including free electrons, and hence the dust particles may also be charged. Therefore, calculating the binding energies of molecules on such charged surfaces is relevant for understanding the properties and evolution of molecular clouds.

On the charged ASW surface, some configurations of CO interact with the charge, here an electron transfer occurs which may lead to a spontaneous H-abstraction forming an HCO radical. This interaction with the extra electron on the surface, increases the binding energy. On the neutral surface, the binding energy of CO lies between 1205 K and 2144 K. The binding energy on a charged surface lies between 1277 K and 5734 K, showing a clear increase in maximum binding energy.

For CH$_4$, there is no significant change in binding energy between the neutral and charged ASW surface. The neutral surface binding energy lies between 920 K and 1771 K, which is broader than the charged range of 1071 K to 1564 K. There is also no interaction observed between CH$_4$ and the surface charge.

Hydrogen bonding plays an important role in the gas-surface interaction between NH$_3$ and the ASW surface. When a charge is added to the surface, we see a decrease in the average binding energy of NH$_3$ on the surface. Adding a charge reduces the binding energy from a range of 4308 K - 14768 K (neutral surface) to a range of 4479 K - 7769 K (charged surface).

These results point to the importance of including surface charge in assessing the reactivity of ASW in the interstellar medium.

\section{Acknowledgments}
\begin{acknowledgments}
The resources and services used in this work were provided by the VSC (Flemish Supercomputer Center), funded by the Research Foundation - Flanders (FWO) and the Flemish Government.
\end{acknowledgments}

\bibliography{References}{}

\begin{thebibliography}{}
\expandafter\ifx\csname natexlab\endcsname\relax\def\natexlab#1{#1}\fi
\providecommand{\url}[1]{\href{#1}{#1}}
\providecommand{\dodoi}[1]{doi:~\href{http://doi.org/#1}{\nolinkurl{#1}}}
\providecommand{\doeprint}[1]{\href{http://ascl.net/#1}{\nolinkurl{http://ascl.net/#1}}}
\providecommand{\doarXiv}[1]{\href{https://arxiv.org/abs/#1}{\nolinkurl{https://arxiv.org/abs/#1}}}

\bibitem[{Adamo \& Barone(1999)}]{Adamo1999}
Adamo, C., \& Barone, V. 1999, The Journal of chemical physics, 110, 6158

\bibitem[{Al-Halabi {et~al.}(2004)Al-Halabi, Fraser, Kroes, \& Dishoeck}]{Al-Halabi2004}
Al-Halabi, A., Fraser, H.~J., Kroes, G.~J., \& Dishoeck, E. F.~V. 2004, Astronomy \& Astrophysics, 422, 777

\bibitem[{Amati {et~al.}(2019)Amati, Stoia, \& Baerends}]{amati2019}
Amati, M., Stoia, S., \& Baerends, E. 2019, Journal of chemical theory and computation, 16, 443

\bibitem[{Autschbach \& Srebro(2014)}]{autschbach2014}
Autschbach, J., \& Srebro, M. 2014, Accounts of chemical research, 47, 2592

\bibitem[{Bovolenta {et~al.}(2022)Bovolenta, Vogt-Geisse, Bovino, \& Grassi}]{bovolenta2022}
Bovolenta, G.~M., Vogt-Geisse, S., Bovino, S., \& Grassi, T. 2022, The Astrophysical Journal Supplement Series, 262, 17

\bibitem[{Cuppen {et~al.}(2017)Cuppen, Walsh, Lamberts, Semenov, Garrod, Penteado, \& Ioppolo}]{cuppen2017}
Cuppen, H., Walsh, C., Lamberts, T., {et~al.} 2017, Space Science Reviews, 212, 1

\bibitem[{Das {et~al.}(2018)Das, Sil, Gorai, Chakrabarti, \& Loison}]{Das2018}
Das, A., Sil, M., Gorai, P., Chakrabarti, S.~K., \& Loison, J.-C. 2018, The Astrophysical Journal Supplement Series, 237, 9

\bibitem[{Draine \& Sutin(1987)}]{draine1987}
Draine, B., \& Sutin, B. 1987, The Astrophysical Journal, 320, 803

\bibitem[{Duflot {et~al.}(2021)Duflot, Toubin, \& Monnerville}]{Duflot2021}
Duflot, D., Toubin, C., \& Monnerville, M. 2021, Frontiers in Astronomy and Space Sciences, 24

\bibitem[{Enrique-Romero {et~al.}(2019)Enrique-Romero, Rimola, Ceccarelli, Ugliengo, Balucani, \& Skouteris}]{Enrique-Romero2019}
Enrique-Romero, J., Rimola, A., Ceccarelli, C., {et~al.} 2019, ACS Earth and Space Chemistry, 3, 2158

\bibitem[{Ferrero {et~al.}(2020)Ferrero, Zamirri, Ceccarelli, Witzel, Rimola, \& Ugliengo}]{Ferrero2020}
Ferrero, S., Zamirri, L., Ceccarelli, C., {et~al.} 2020, The Astrophysical Journal, 904, 11

\bibitem[{Frisch {et~al.}(2016)Frisch, Trucks, Schlegel, Scuseria, Robb, Cheeseman, Scalmani, Barone, Petersson, Nakatsuji, Li, Caricato, Marenich, Bloino, Janesko, Gomperts, Mennucci, Hratchian, Ortiz, Izmaylov, Sonnenberg, Williams-Young, Ding, Lipparini, Egidi, Goings, Peng, Petrone, Henderson, Ranasinghe, Zakrzewski, Gao, Rega, Zheng, Liang, Hada, Ehara, Toyota, Fukuda, Hasegawa, Ishida, Nakajima, Honda, Kitao, Nakai, Vreven, Throssell, Montgomery, Peralta, Ogliaro, Bearpark, Heyd, Brothers, Kudin, Staroverov, Keith, Kobayashi, Normand, Raghavachari, Rendell, Burant, Iyengar, Tomasi, Cossi, Millam, Klene, Adamo, Cammi, Ochterski, Martin, Morokuma, Farkas, Foresman, \& Fox}]{g16}
Frisch, M.~J., Trucks, G.~W., Schlegel, H.~B., {et~al.} 2016, Gaussian˜16 {R}evision {C}.01

\bibitem[{Grimme {et~al.}(2011)Grimme, Ehrlich, \& Goerigk}]{Grimme2011}
Grimme, S., Ehrlich, S., \& Goerigk, L. 2011, Journal of computational chemistry, 32, 1456

\bibitem[{Ivlev {et~al.}(2015)Ivlev, Padovani, Galli, \& Caselli}]{ivlev2015}
Ivlev, A.~V., Padovani, M., Galli, D., \& Caselli, P. 2015, The Astrophysical Journal, 812, 135

\bibitem[{Jensen(2010)}]{jensen2010}
Jensen, F. 2010, Journal of chemical theory and computation, 6, 2726

\bibitem[{Martínez {et~al.}(2009)Martínez, Andrade, Birgin, \& Martínez}]{Martinez2009}
Martínez, L., Andrade, R., Birgin, E.~G., \& Martínez, J.~M. 2009, Journal of computational chemistry, 30, 2157

\bibitem[{Mestel \& Spitzer~Jr(1956)}]{mestel1956}
Mestel, L., \& Spitzer~Jr, L. 1956, Monthly Notices of the Royal Astronomical Society, 116, 503

\bibitem[{Minissale {et~al.}(2022)Minissale, Aikawa, Bergin, Bertin, Brown, Cazaux, Charnley, Coutens, Cuppen, \& Guzman}]{Minissale2022}
Minissale, M., Aikawa, Y., Bergin, E., {et~al.} 2022, ACS Earth and Space Chemistry, 6, 597

\bibitem[{Okuzumi(2009)}]{okuzumi2009}
Okuzumi, S. 2009, The Astrophysical Journal, 698, 1122

\bibitem[{Pantaleone {et~al.}(2020)Pantaleone, Enrique-Romero, Ceccarelli, Ugliengo, Balucani, \& Rimola}]{Pantaleone2020}
Pantaleone, S., Enrique-Romero, J., Ceccarelli, C., {et~al.} 2020, The Astrophysical Journal, 897, 56

\bibitem[{Penteado {et~al.}(2017)Penteado, Walsh, \& Cuppen}]{penteado2017}
Penteado, E., Walsh, C., \& Cuppen, H. 2017, The Astrophysical Journal, 844, 71

\bibitem[{Perdew {et~al.}(1996)Perdew, Ernzerhof, \& Burke}]{Perdew1996}
Perdew, J.~P., Ernzerhof, M., \& Burke, K. 1996, The Journal of chemical physics, 105, 9982

\bibitem[{Perdew {et~al.}(1982)Perdew, Parr, Levy, \& Balduz~Jr}]{perdew1982}
Perdew, J.~P., Parr, R.~G., Levy, M., \& Balduz~Jr, J.~L. 1982, Physical Review Letters, 49, 1691

\bibitem[{Perdew \& Zunger(1981)}]{perdew1981}
Perdew, J.~P., \& Zunger, A. 1981, Physical Review B, 23, 5048

\bibitem[{Rimola {et~al.}(2021)Rimola, Ceccarelli, Balucani, \& Ugliengo}]{Rimola2021}
Rimola, A., Ceccarelli, C., Balucani, N., \& Ugliengo, P. 2021, Frontiers in Astronomy and Space Sciences, 8, 655405

\bibitem[{Rimola {et~al.}(2014)Rimola, Taquet, Ugliengo, Balucani, \& Ceccarelli}]{Rimola2014}
Rimola, A., Taquet, V., Ugliengo, P., Balucani, N., \& Ceccarelli, C. 2014, Astronomy \& Astrophysics, 572, A70

\bibitem[{Spitzer~Jr(1941)}]{spitzer1941}
Spitzer~Jr, L. 1941, Astrophysical Journal, vol. 93, p. 369, 93, 369

\bibitem[{Tielens(2005)}]{Tielens2005}
Tielens, A. G. G.~M. 2005, The physics and chemistry of the interstellar medium (Cambridge University Press)

\bibitem[{Tinacci {et~al.}(2022)Tinacci, Germain, Pantaleone, Ferrero, Ceccarelli, \& Ugliengo}]{Tinacci2022}
Tinacci, L., Germain, A., Pantaleone, S., {et~al.} 2022, ACS Earth and Space Chemistry, 6, 1514, \dodoi{10.1021/acsearthspacechem.2c00040}

\bibitem[{Tsuge \& Watanabe(2021)}]{Tsuge2021}
Tsuge, M., \& Watanabe, N. 2021, Accounts of Chemical Research, 54, 471, \dodoi{10.1021/acs.accounts.0c00634}

\bibitem[{Vydrov {et~al.}(2007)Vydrov, Scuseria, \& Perdew}]{vydrov2007}
Vydrov, O.~A., Scuseria, G.~E., \& Perdew, J.~P. 2007, The Journal of chemical physics, 126

\bibitem[{Wakelam {et~al.}(2017)Wakelam, Loison, Mereau, \& Ruaud}]{Wakelam2017}
Wakelam, V., Loison, J.-C., Mereau, R., \& Ruaud, M. 2017, Molecular Astrophysics, 6, 22

\bibitem[{Weigend(2006)}]{Weigend2006}
Weigend, F. 2006, Phys. Chem. Chem. Phys., 8, 1057, \dodoi{10.1039/B515623H}

\bibitem[{Woon(2023)}]{Woon2023}
Woon, D.~E. 2023, Monthly Notices of the Royal Astronomical Society, 527, 1357, \dodoi{10.1093/mnras/stad3242}

\bibitem[{Zamirri {et~al.}(2018)Zamirri, Casassa, Rimola, Segado-Centellas, Ceccarelli, \& Ugliengo}]{zamirri2018}
Zamirri, L., Casassa, S., Rimola, A., {et~al.} 2018, Monthly Notices of the Royal Astronomical Society, 480, 1427

\end{thebibliography}
\bibliographystyle{aasjournal}

\end{document}